# A comparison of apartment rent price prediction using a large dataset: Kriging versus DNN

Hajime Seya
Departments of Civil Engineering, Graduate School of Engineering Faculty of Engineering, Kobe University, 1-1, Rokkodai, Nada, Kobe, Hyogo, 657-8501, Japan,
E-mail: hseya@people.kobe-u.ac.jp; Tel.: +81-78-803-6278

Daiki Shiroi
Gunosy Inc., Roppongi Hills Mori Tower, 6-10-1, Roppongi, Minato, Tokyo, E-mail: daikishiroi@gmail.com

***Abstract***

The hedonic approach based on a regression model has been widely adopted for the prediction of real estate property price and rent. In particular, a spatial regression technique called Kriging, a method of interpolation that was advanced in the field of spatial statistics, are known to enable high accuracy prediction in light of the spatial dependence of real estate property data. Meanwhile, there has been a rapid increase in machine learning-based prediction using a large (big) dataset and its effectiveness has been demonstrated in previous studies. However, no studies have ever shown the extent to which predictive accuracy differs for Kriging and machine learning techniques using big data. Thus, this study compares the predictive accuracy of apartment rent price in Japan between the nearest neighbor Gaussian processes (NNGP) model, which enables application of Kriging to big data, and the deep neural network (DNN), a representative machine learning technique, with a particular focus on the data sample size ($n$ = $10^4$, $10^5$, $10^6$) and differences in predictive performance. Our analysis showed that, with an increase in sample size, the out-of-sample predictive accuracy of DNN approached that of NNGP and they were nearly equal on the order of $n = 10^6$. Furthermore, it is suggested that, for both higher and lower end properties whose rent price deviates from the median, DNN may have a higher predictive accuracy than that of NNGP.

***Acknowledgment***

This study was supported by the Grants-in-Aid for Scientific Research Grant No. 18H03628 and 17K14738 from the Japan Society for the Promotion of Science. Furthermore, the "LIFULL HOME'S Data Set," was provided for research purposes by LIFULL Co., Ltd. with the assistance of the National Institute of Informatics, Japan.

# A comparison of apartment rent price prediction using a large dataset: Kriging versus DNN

## 1. Introduction

Real estate is an industry that has been said to relatively lag behind other businesses in terms of digitalization. During recent years, however, efforts to streamline operations using technologies have been gaining momentum. Online automated services for property price estimation is a recent technology and Zillow[1], a service offered by Zillow Group in the US, is well known. Likewise, in Japan, there is a service termed Price Map[2] from LIFULL. In Price Map, for example, properties are represented on a map with which one can review reference sale and rent prices by entering information such as room layout and lot size. Other similar services also exist and they are typically supported by huge property databases and statistics- or machine-learning-based sale and rent price prediction technologies. Against the backdrop of improved computer performance and expanded databases, these technologies have been termed as big data analysis or AI during recent years, and they are making rapid progress. As exemplified by the term *ReTech*, this has undoubtedly significantly impacted the real estate industry.

Behind the increase in automated price assessment and prediction technologies using big data is the inefficiency of the current real estate industry. For example, in Japan, licensed real estate appraisers provide a property assessment based on an expected cash flows and comparison with similar properties. However, because all real estate properties have a *uniqueness* in the sense that there are no two identical properties and because an appraisal must be conducted considering the supply/demand balance, in addition to the property characteristics themselves, a tremendous effort must be made in explaining for the basis for the appraisal to consumers. Aside from appraisal costs, there is an issue of information asymmetry between real estate agencies and general consumers as seen in the limited disclosure of purchase prices because of privacy policy[3]. This means that if real estate sale or rent prices can be reasonably and quickly predicted, then the information asymmetry between consumers and real estate agencies would disappear, leading to revitalization of the real-estate market. In other words, there is a need for an accurate prediction model of real estate sale and rent prices for businesses and consumers.

[1] https://www.zillow.com/
[2] https://www.homes.co.jp/price-map

While attention has been drawn to automated assessment of real-estate sale and rent prices using big data and machine learning techniques (Abidoye and Chan, 2017), conventionally, the hedonic approach has been widely used for predicting real estate sale and rent prices (Rosen, 1974). The hedonic approach takes the real estate sale or rent price to be a sum total of the values of attributes that comprise a property and typically estimates its price through regression analysis. Because it is an approach based on regression analysis, it can be implemented with relative ease and more importantly, marginal benefits of attributes can be evaluated using a calculated regression coefficient as a by-product. Meanwhile, from the perspective of prediction, simple functional forms such as logarithmic form or Box-Cox form are typically used[4]. Thus, in a context where the sample size is sufficiently large, the big data plus machine learning approach, which can construct complex non-linear functions, is expected to outperform its alternatives. Indeed, as mentioned in the next section, several studies show this trend.

For real estate appraisal, there are factors that are difficult to accommodate as explanatory variables such as local brand and historical context. It is therefore important to consider how these unobserved factors can be incorporated into the model. In spatial statistics, a method of handling these variables as the spatial dependence of error terms (neighboring properties are prone to a similar error), more strictly, a method of capturing spatial dependence by assuming a Gaussian process (GP), was established as Kriging (Dubin, 1988). To name only a few, James et al. (2005), Bourassa et al. (2010), and Seya et al. (2011) have reported that Kriging provides high predictive accuracy compared to simple multiple regression models (hereinafter referred to as ordinary least squares (OLS)). Because model structure of OLS is rather simple, parameters can be properly set even with a relatively small sample size and there are not many benefits in using big data. By contrast, with Kriging, because pricing information of neighboring properties is reflected in the predicted results through spatial dependence, the situation is different from that of OLS.

[3]For example, although information on individual transactions of real estate properties is officially available in the Land General Information System (http://www.land.mlit.go.jp/webland/) from the Ministry of Land, Infrastructure, Transport and Tourism in Japan, location and price details are not available, and only approximate locations and prices are available.
[4] Semiparametric functional forms such as penalized spline can also be used (Seya et al., 2011).

As previously noted, because the regression-based model can be used for evaluation, it is of high practical use particularly in the social sciences. It is therefore important to test the extent to which its predictive accuracy differs from that of the machine learning approach and understand the order.[5] Thus, the aim of this study was to compare and discuss the results of rent price prediction using three different approaches—the [1] OLS, [2] spatial statistical (Kriging), and [3] machine learning approaches—for various sample sizes. As the sample size increases, it is increasingly more difficult to straightforwardly apply Kriging which requires the cost of $O(n^3)$ for inverting a variance–covariance matrix (for example, when $n = 10^5$). Hence as a spatial statistical approach, the nearest neighbor Gaussian processes (NNGP) model was used, which allows application of Kriging to big data (Datta et al., 2016; Finley et al., 2017; Zhang et al., 2019). While there are various approaches for spatial statistical modeling using big data, NNGP has been demonstrated to have better predictive accuracy and practicality in a comparative study (Heaton et al., 2018).

As a machine learning approach, the deep neural network (DNN) was used. Neural networks including the DNN are mathematical models of the information processing mechanism in the brain composed of billions of neurons; multi-layered neural networks are termed the DNN. The DNN can construct highly complicated non-linear functions and can consider spatial dependence through a non-linear function for position coordinates without explicitly modeling the spatial dependence as in the spatial statistical approach (e.g., Cressie and Wikle, 2011).

For validation, "LIFULL HOME'S Data Set"[6], a data set for apartment rent prices in Japan provided by LIFULL Co., Ltd. free of cost through the National Institute of Informatics, to researchers, was used. Because there is generally less research regarding rental price than that regarding sales price, it may be a valuable data set. The data set consists of snapshots (cross-section data) that are either rental property data or image data as of September 2015. The former shows rent, lot size, location (municipality, zip code, nearest station, and walk time to nearest station), year built, room layout, building structure, and equipment for 5.33 million properties throughout Japan whereas the latter is comprised of 83 million pictures that show the floor plan and interior details for each property. In this study, only the former was used. One of our future tasks is to perform an experiment using the latter.

[5] There are models such as regression trees that are based on machine learning that can nevertheless provide useful interpretations.

[6] https://www.nii.ac.jp/dsc/idr/lifull/homes.html

Out of approximately 5.33 million properties, 4,588,632 properties were obtained after excluding missing data, from which $n = 10^4$, $10^5$, and $10^6$ properties were randomly sampled. While focusing on the difference in sample size, the accuracies of out-of-sample prediction for property rent price based on approaches [1], [2], and [3] were compared through validation. The number of explanatory variables $K$ was 43 including constant terms. Our analysis showed that with an increase in sample size, the predictive accuracy of DNN was observed to approach that of NNGP and on the order of $n = 10^6$ they were nearly equal. During this experiment, standard explanatory variables that had been incorporated into the regression-based hedonic model were used. Our findings suggested that, using these standard settings, even if the sample size is on the order of $n = 10^6$, the use of regression-based NNGP is sufficient.

In Chapter 2, we briefly review previous studies regarding related issues. Chapter 3 briefly explains models used in this comparison study. Chapter 4 shows the results of the comparative analysis using the LIFULL HOME'S data set. Lastly, Chapter 5 presents conclusions and provides future challenges to address.

## 2. Literature review

This chapter offers a review of previous studies that compared the regression-based approach and the neural-network-based approach in terms of prediction of real estate sale and rent prices. Against, perhaps, readers' anticipation, relatively limited research is available on this topic.

Kontrimas and Verikas (2011) compared the predictive accuracy of the machine-learning approach including multi-layer perceptron (MLP), a subset of DNN, and OLS using data on home sale transactions. They found that the mean absolute percentage difference (MAPD) for MLP and OLS was 23% and 15%, respectively, and MLP was outperformed by OLS. However, the sample size of their study was no greater than 100. Similarly, Georgiadis (2018) compared the predictive accuracies of regression-based models and Artificial Neural Networks (ANN) on sales prices of 752 apartments in Thessaloniki, Greece, using cross valuation and found that the geographically weighted regression model (Fotheringham et al., 2002) outperformed ANN. While these two studies have shown that the regression-based approach outperformed the neural-network-based approach in terms of predictive accuracy, the sample sizes used for these studies were merely on the order of $n = 10^2$.

Meanwhile, Abidoye and Chan (2018) compared ANN and OLS using sales transaction data for 321 residential properties in Lagos, Nigeria, and concluded that ANN outperformed OLS. Likewise, Yalpır (2018) and Selim (2009) compared ANN and OLS and suggested that the former performed better. Yalpır (2018) used 98 study samples, whereas Selim (2009) used fairly large—5741 samples. In Yalpır (2018), they used three activation functions (the sigmoid, tangent hyperbolic, and adaptive activation functions) to build ANN. However, hyperparameters other than activation functions were fixed in validation.

As previously discussed, although several attempts have been made to compare and examine the predictive accuracy of real estate sale and rent prices between a regression-based approach and a neural-network-based approach, the results obtained were largely mixed. Limitations of previous studies include [1] a small sample size (except for Selim (2009)), [2] disregard of spatial dependence which is an essential characteristic of real-estate properties (except for Georgiadis (2018)), and [3] tailored and ad hoc settings of hyperparameters in DNN (or ANN). To address these challenges, the present study attempted to [1] perform an experiment at different and relatively large-scale sample sizes ($n = 10^4$, $10^5$, $10^6$), [2] consider the spatial dependence either the application of NNGP (Kriging) or the function of latitude/longitude coordinates (in case of DNN), and [3] optimize hyperparameters in DNN.

## 3. Model

### 3.1. Nearest Neighbor Gaussian Processes (NNGP)

Let $D$ be the spatial domain under study and let $\boldsymbol{s}$ be a coordinate position (x, y coordinates). Then, the spatial regression model, often termed as the spatial process model, can be expressed as follows:

$$y(\boldsymbol{s}) = m(\boldsymbol{s}) + w(\boldsymbol{s}) + \varepsilon(\boldsymbol{s}), \qquad \varepsilon(\boldsymbol{s}) \sim N(0, \tau^2), \tag{1}$$

where $\tau^2$ is a variance parameter termed a nugget that represents micro scale variation and/or measurement error. Normally, we assume that $m(\boldsymbol{s}) = \boldsymbol{x}(\boldsymbol{s})'\boldsymbol{\beta}$, where $\boldsymbol{x}$ is an explanatory parameter vector at the point $\boldsymbol{s}$ and $\boldsymbol{\beta}$ is the corresponding regression coefficient vector. $w(\boldsymbol{s})$ is assumed to follow the Gaussian process (GP) $w(\boldsymbol{s}) \sim GP\big(0, C(\cdot,\cdot\,|\boldsymbol{\theta})\big)$, where the mean is zero and the covariance function is $C(\cdot,\cdot\,|\boldsymbol{\theta})$ ($\boldsymbol{\theta}$ is a parameter vector that normally includes the parameter $\phi$ [where $1/\phi$ is called the range], which controls the range of spatial dependence, and the variance parameter $\sigma^2$, which represents the variance of spatial process and termed partial sill). If a sample is obtained at point $\boldsymbol{s}_1, \ldots, \boldsymbol{s}_n$, $\boldsymbol{w} = \big(w(\boldsymbol{s}_1), w(\boldsymbol{s}_2), \ldots, w(\boldsymbol{s}_n)\big)'$ follows the multivariate Gaussian distribution : $\boldsymbol{w} \sim N(\boldsymbol{0}, \boldsymbol{C}(\boldsymbol{\theta}))$, where the mean is zero and the covariance function is $C(\boldsymbol{s}_i, \boldsymbol{s}_j|\boldsymbol{\theta})$. Then the spatial regression model can be expressed as $\boldsymbol{y} \sim N(\boldsymbol{X\beta},\ \boldsymbol{\Lambda}(\tau^2, \boldsymbol{\theta}))$, where $\boldsymbol{\Lambda}(\boldsymbol{\theta}) = \boldsymbol{C}(\boldsymbol{\theta}) + \tau^2\boldsymbol{I}$ with $\boldsymbol{I}$ is an $n \times n$ identity matrix. If the following relationship holds for any movement $\boldsymbol{h} \in D$, $w(\boldsymbol{s})$ is considered a second-order stationarity spatial process.

$$E[w(\boldsymbol{s})] = 0; \forall \boldsymbol{s} \in D, \tag{2}$$

$$Cov[w(\boldsymbol{s}), w(\boldsymbol{s}+\boldsymbol{h})] = C(\boldsymbol{h}|\boldsymbol{\theta});\ \forall \boldsymbol{s}, \boldsymbol{h} \in D, \tag{3}$$

$$Cov[w(\boldsymbol{s}), w(\boldsymbol{s}+\boldsymbol{0})] = Var[w(\boldsymbol{s})] = C(\boldsymbol{0}|\boldsymbol{\theta});\ \forall \boldsymbol{s}, \boldsymbol{h} \in D. \tag{4}$$

The second-order stationarity assumes that the covariance does not depend on position $\boldsymbol{s}$ and only on $\boldsymbol{h}$. When $\boldsymbol{h}$ depends only on the distance $d = ||\boldsymbol{h}||$ and not the direction, the spatial process is said to have isotropy (|| • || is the vector norm). Covariance functions $C(\boldsymbol{s}_i, \boldsymbol{s}_j|\boldsymbol{\theta})$ that meet second-order stationarity can be spherical, Gaussian, exponential, Matérn, etc. (Cressie, 1993).

The prediction of $y(\boldsymbol{s_0})$ at any given point $\boldsymbol{s_0}$ is termed Kriging[7]. The Kriging predictor includes the $n \times n$ variance-covariance matrix $\boldsymbol{C}$ with elements represented by $C(\boldsymbol{s}_i, \boldsymbol{s}_j|\boldsymbol{\theta})$ and the calculation requires the cost of $O(n^3)$. The calculation will be difficult when $n$ takes the value around $n = 10^5$. By contrast, various approaches have been proposed that approximate the spatial process $\boldsymbol{w}(\boldsymbol{s})$ (Heaton et al., 2018). Among other alternatives, this study used the NNGP model. NNGP is based on Vecchia (1988) and assumes the following approximation to the joint likelihood $p(\boldsymbol{w}) = p(w(\boldsymbol{s}_1)) \prod_{i=2}^{n} p(w(\boldsymbol{s}_i)|w(\boldsymbol{s}_1), w(\boldsymbol{s}_2) \ldots, w(\boldsymbol{s}_{i-1}))$.[8]

$$\tilde{p}(\boldsymbol{w}) = p(w(\boldsymbol{s}_1)) \prod_{i=2}^{n-1} p(w(\boldsymbol{s}_i)|\boldsymbol{w}(N(\boldsymbol{s}_i))). \tag{5}$$

[7] Or $m(\boldsymbol{s_0}) + w(\boldsymbol{s_0})$, see Cressie (1993).

[8] Although the results depend on the ordering of the samples, Datta et al. (2016) showed that NNGP is insensitive to ordering. We performed ordering based on the x-coordinate locations.

Here, $N(\boldsymbol{s}_i)$ is a neighbors set of $\boldsymbol{s}_i$, and it is given as the *k*-nearest neighbors of $\boldsymbol{s}_i$ in NNGP. Thus, NNGP approximates the full GP expressed as a joint distribution using the nearest neighbors. Datta et al. (2016) demonstrated that the approximation of formula (5) leads to approximation of the precision matrix $\boldsymbol{C}^{-1}$ to $\widetilde{\boldsymbol{C}}^{-1}$ provided in the following formula:

$$\widetilde{\boldsymbol{C}}^{-1} = (\boldsymbol{I} - \boldsymbol{A})'\boldsymbol{D}(\boldsymbol{I} - \boldsymbol{A}) \tag{6}$$

where $\boldsymbol{A}$ is a sparse lower triangular matrix with at most *k*-entries in each row and $\boldsymbol{D}$ = diag($d_{ii}$) is a diagonal matrix. Here, because $\boldsymbol{A}$ and $\boldsymbol{D}$ can be provided as $m \times m$ matrices, it can significantly reduce the computational load. The spatial regression model provided through NNGP may be written as follows:

$$\boldsymbol{y} \sim N(\boldsymbol{X\beta},\ \widetilde{\boldsymbol{\Lambda}}(\tau^2, \boldsymbol{\theta})) \tag{7}$$

where $\widetilde{\boldsymbol{\Lambda}}(\tau^2, \boldsymbol{\theta}) = \widetilde{\boldsymbol{C}}(\boldsymbol{\theta}) + \tau^2\boldsymbol{I}$.

The NNGP model parameters can be estimated using (Bayesian) Markov Chain Monte Carlo (MCMC) (Datta et al., 2016), Hamiltonian Monte Carlo (Wang et al., 2018), and maximum likelihood methods (Saha and Datta, 2018). This study uses MCMC. Because the NNGP parameters are $\boldsymbol{\beta}$ and $\boldsymbol{\phi} = (\tau^2, \sigma^2, \phi)' = (\tau^2, \boldsymbol{\theta})'$, when using MCMC, we need to set a prior distribution for each parameter and multiply it by the likelihood function to obtain conditional posterior distributions (full Bayesian NNGP). Because this study addresses massive data up to a maximum of $n = 10^6$, it is difficult to implement the full Bayesian NNGP within a practical computational time. Accordingly, conjugate NNGP as proposed by Finley et al. (2017) was used. Suppose $\widetilde{\boldsymbol{P}}(\phi)$ is a nearest neighbors approximation of a correlational matrix corresponding to a nearest neighbors approximation of a variance-covariance matrix—$\widetilde{\boldsymbol{C}}(\boldsymbol{\theta})$. Then the conjugate NNGP can be provided as follows:

$$\boldsymbol{y} \sim N(\boldsymbol{X\beta},\ \sigma^2\widetilde{\boldsymbol{M}}) \tag{8}$$

where $\widetilde{\boldsymbol{M}} = \widetilde{\boldsymbol{P}}(\phi) + \alpha\boldsymbol{I}$ and $\alpha = \tau^2/\sigma^2$. The point of using the conjugate NNGP is that, when assuming that $\alpha$ and $\phi$ are known, the conjugate normal-inverse Gamma posterior distribution for $\boldsymbol{\beta}$ and $\sigma^2$ can be used and the predictive distribution for $y(\boldsymbol{s_0})$ can also be obtained as a *t*-distribution; thus, it is extremely easy to perform sampling.

### 3.2. Deep neural network (DNN)

DNN is inspired by organism's neural networks. It is a mathematical model that has a network structure in which layered units are connected with neighboring layers. DNN allows construction of extremely complicated non-linear functions. What follows is a schematic diagram of a standard three-layered DNN created in reference to Raju et al. (2011):

[Figure 1: Three-layered feedforward neural network

(Created by author in reference to Raju et al. (2011))], around here

Each element that comprises a network is termed a unit or node and is represented as O (circle) in Figure 1. The first layer is termed the input layer and the last the output; all of the other layers are referred to as hidden layers. In DNNs, results of non-linear transformations on inputs received from the previous layer are transmitted to the next layer to ultimately derive a single output as an estimation result. In doing so, linear transformations via a weighted matrix $\boldsymbol{W}_l$ $(m_l \times m_{(l+1)})$ and non-linear transformations via an activation function *f*(.) occur in each layer. The transformation from the $l^{th}$ layer output $z_l$ $(m_l \times 1)$ to the $l + 1^{th}$ layer output $z_{l+1}$ $(m_{(l+1)} \times 1)$ can be computed according to the following formulas:

$$\boldsymbol{u}_{l+1} = \boldsymbol{W}_l \boldsymbol{z}_l + \boldsymbol{b}_{l+1} , \tag{9}$$

$$\boldsymbol{z}_{l+1} = f(\boldsymbol{u}_{l+1}). \tag{10}$$

where $\boldsymbol{b}$ is a bias term. Suppose the number of layers is expressed as $l = 1, \ldots, L$, output $\boldsymbol{z}_L$ in the $L^{th}$ layer is the final output $(\boldsymbol{z}_L \equiv \hat{y})$. *f*(.) is a non-linear function termed an activation function. Typical activation functions include the sigmoid and Rectified Linear Unit, Rectifier (ReLU) functions; the latter was used in this study. Any differentiable function can be used for the loss function of a DNN. If regression is used as in this study, the Mean Squared Error (MSE) of the actual value $y$ and the predictive value $\hat{y}$ is often used.

$$h = \frac{1}{n}\sum_{i=1}^{n}(y_i - \hat{y}_i)^2 \tag{11}$$

For the loss function $h$, searching $\boldsymbol{W}$ and $\boldsymbol{b}$ that minimize $h$ is termed DNN learning. Learning is performed by the gradient algorithm, while backpropagation is used to calculate the gradient. In contrast to the case of estimation, partial derivatives are computed in order from the output layer (LeCun et al., 2011).

## 4. Comparison experiments: Kriging versus DNN

### 4.1. Dataset

As previously mentioned in Chapter 1, the LIFULL HOME'S data set was used in this study for rent price prediction. Out of approximately 5.33 million properties, 4,588,632 properties obtained by excluding missing data were used as original data. Although the original data did not explicitly contain property positional coordinates s, they did contain zip codes and barycentric coordinates for zip codes (X,Y coordinates of a WGS84 UTM54N type) were used on their behalf. This led to inclusion of some positional errors in the positional coordinates. However, given that our study was nationwide in scope, these errors are ignorable. The dependent variable is the natural logarithm of the rent price (yen, including maintenance fees)) the explanatory variables shown in Table 1 were used. The number of explanatory variables (*K*) was 43. Table 1 also shows descriptive statistics. Although the classification of room layout in Table 1 could be more fine-grained, we used a slightly coarse classification because we were more interested in comparing models than building a perfect hedonic model. Of all the explanatory variables, information regarding use district (zoning) and floor-area ratio was often lacking in the original database. Therefore, these data were separately prepared from the National Land Numerical

Information[9]. Figure 2 shows the number of properties per 1000 $km^2$ and Figure 3 shows the natural logarithm of the rent price (yen) for each prefecture.

[Table 1-1: Descriptive statistics (Continuous variables)], around here

[Table 1-2: List of explanatory variables (Discrete variables)], around here

[Table 1-3: Descriptive statistics (Discrete variables)], around here

[Figure 2: Number of properties per 1000 $km^2$ for each prefecture], around here

[Figure 3: log (rent price) for each prefecture], around here

### 4.2. Experimental design

We compared the rent prediction accuracy based on three models: OLS, NNGP, and DNN. For prediction, of 4,588,632 properties, properties were randomly selected at various sizes ($n = 10^4$, $10^5$, $10^6$) and 80% of these data were used as training data for models for learning and the remaining 20% were used as testing data (validation data) to test the prediction accuracy. The sample size for training and testing data had three patterns: (8000 vs. 2000), (80,000 vs. 20,000), and (800,000 vs. 200,000). Because sampling was completely randomly conducted, there were no containment relations such that, for example, $10^4$ samples are contained in $10^5$ samples. However, because the data size was sufficiently big, it would be highly unlikely that the sample bias would conceal trends, and thus this study design (based not on conditionalization but on complete random sampling) would not greatly affect results.

For predictive accuracy assessment, the following error measures were used. Here, $\hat{y}_i$, $y_i$ are the out-of-sample predictive and observed values, respectively, for the $i^{th}$ data.

$$MAE = \frac{1}{n}\sum_{i=1}^{n}|y_i - \hat{y}_i| \tag{12}$$

$$MSE = \frac{1}{n}\sum_{i=1}^{n}(y_i - \hat{y}_i)^2 \tag{13}$$

$$RMSE = \sqrt{\frac{1}{n}\sum_{i=1}^{n}(y_i - \hat{y}_i)^2} \tag{14}$$

[9] http://nlftp.mlit.go.jp/ksj-e/index.html

$$MAPE = \frac{100}{n}\sum_{i=1}^{n}\left|\frac{y_i - \hat{y}_i}{y_i}\right| \quad (15)$$

## 4.3. Model settings

### 4.3.1. OLS

OLS was added for comparison as a usual hedonic regression model that does not consider spatial dependence. The explanatory variables are shown in Table 1 save the X and Y coordinates. As a reference, Table 2 shows regression analysis results based on the OLS estimation when $n = 10^6$. The adjusted $R^2$ value was 0.5178 and fairly good given the sample size.

[Table 2: Regression analysis results using OLS (example of $n = 10^6$)], around here

### 4.3.2. NNGP

We used the conjugate NNGP proposed by Finley et al. (2017) as explained in Section 3.1. The conjugate NNGP is a pragmatic approach that accelerates sampling by assuming $\alpha$ and $\phi$ to be "known." Needless to say, the full Bayesian NNGP is theoretically sound. In this study, however, we addressed massive data with up to $n = 10^6$ of data; hence, it is practically difficult to implement full Bayesian NNGP. In cases such as this, the conjugate NNGP offers a very useful alternative. Finley et al. (2017) proposed to assign values to $\alpha$ and $\phi$ via the grid point search algorithm based on the cross-validation (CV) score. However, the computational load is high for performing a grid point search for $n = 10^6$ of data. Therefore, in this study, the following simplified procedure was undertaken in assigning values to $\alpha$ and $\phi$[10]. From the remaining data that were not used for comparison in this study, 10,000 properties were randomly sampled and parameters were defined by iteratively re-weighted generalized least squares (Schabenberger and Gotway, 2005, pp. 256–259) in the semivariogram $\gamma(\mathbf{h}) = C(\mathbf{0}) - C(\mathbf{h})$, which is in converse relation to the covariance function. Figure 4 shows the fitting results. Starting from the left, the Gaussian, spherical, and exponential models are shown. Of these, the Gaussian model had the best CV score, and hence, was used. We can see that the Gaussian model is a particularly good fit to near-distance that is subject to prediction results. Given these observations, the value for each parameter was as follows: $\phi$=1/25.8, $\tau^2$ = 0.04, and $\sigma^2$ = 0.03.

[10] One possible means to improve this is to apply the methods of hyper parameters value setting for the DNN as mentioned in the next section. For the development of a concrete algorithm, we are leaving it for future study.

[Figure 4: Fitting of variogram functions

(Gaussian model; Spherical model; Exponential model)], around here

As a next step, the model parameters thus created were used to develop an NNGP model. For implementation, the spConjNNGP function in the spNNGP package of R was used. An NNGP model requires determining the number of nearest neighbors to consider. In the default setting of the spConjNNGP function, it is 15[11]. When the relation between the number of nearest neighbors $k$ and CV score (MSE) was plotted[12], there was a tendency for the MSE to decrease to approximately $k$ = 30 and then increase (Figure 5). Thus, the number of nearest neighbors was set as $k$ = 30 in performing the validation.

[Figure 5: Change in the MSE according to the number of nearest neighbors (in the case of n=$10^5$)]

#### 4.3.3. DNN

This subsection explains the DNN settings. DNN has a number of parameters to be determined, including the number of layers, the number of units in the hidden layers, learning rate, and batch size. In addition, the DNN parameter space has a tree structure, which means that we must be aware of the presence of conditional parameters. For example, the number of units in each layer cannot be determined until the determination of the number of layers. The presence of these hyper parameters is undoubtedly a source of the plasticity and high predictive accuracy of a DNN. Conversely, there is no denying that the difficulty in and personalization of settings are obstacles for applied researchers and practitioners who are interested in the prediction of real estate sale and rent prices.

Thus, in this study, optimization for hyperparameters setting were considered. The grid and random searches are widely known as typical methods for DNN parameter tuning (Bergstra and Bengio, 2012). In this study, a more efficient optimization technique known as the tree-structured Parzen estimator (TPE) was adopted (Bergstra et al., 2011). The reason for adoption is its ability to well address the tree-structured parameter space of DNNs and its numerous records of adoption with proven performance to some degree (Bergstra et al., 2011; Bergstra et al., 2013). Nevertheless, the parameter space (range of search) must be given a priori, and after much trial and error, it was set as shown in Table 3.

[11] In the default setting of the spConjNNGP function, the value is 15.
[12] Because $n = 10^4$ and $n = 10^6$ did not produce large differences, the results of $n = 10^5$ are shown here.

[Table 3: DNN hyper parameter and search range], around here

ReLU[13] and MSE (refer to §3.2) were used for the activation function and loss function, respectively. Regarding the optimizer for the DNN, because relatively large differences were found in the results according to the type of algorithm used, results using typical algorithms, RMSprop (Tieleman and Hinton, 2012) and Adam (Adaptive moment estimation) (Kingma and Ba, 2014), are shown. Techniques designed to prevent overtraining such as regularized terms and dropout were not used in this study. Keras[14] was used for the development of a DNN, and Optuna[15], a framework developed via Preferred Networks, Inc., was used for TPE implementation.

The learning procedure for a concrete model was undertaken as follows. First, based on the $t$ th hyper-parameter candidate vectors $\boldsymbol{\delta}_t$ and the results of applying a five-fold cross validation with training data for each $\boldsymbol{\delta}_t$ (MSE, eq. (11)), a 50-fold search was performed using TPE. Second, a model was created once again using the optimal hyper-parameter vector thus obtained and all the training data to assess the predictive accuracy of the testing data. The explanatory variables used were standardized in advance. Table 4 shows the optimization results of the hyper parameters.

[Table 4: DNN hyper parameters after optimization], around here

[13] Historically, Sigmoid and Tanh were primarily used. Currently, ReLU has been accepted as a standard activation function (LeCun et al., 2011).
[14] https://keras.io
[15] https://optuna.org

### 4.4. Results

The predictive accuracies by sample size for each model are shown in Table 5.

[Table 5: Prediction results by sample size for each model], around here

As shown in Table 5, as a DNN optimizer, Adam had considerably higher predictive accuracy compared to that of RMSprop. Therefore, for comparison to other models, Adam was used as a reference. The predictive accuracies of OLS did not display large differences even if the sample size increased. This would be because OLS, which did not use local spatial information, has a simple model structure such that $n$ = $10^4$ was sufficiently large for determining parameters. NNGP demonstrated the best results of all three models, for any sample size and any error measures. Even with a relatively smaller sample size ($n$ = $10^4$), it showed high accuracy (MAPE = 1.152). At $n$ =$10^4$, DNN had a larger error than that of OLS when considering the root mean square error (RMSE) (OLS: RMSE = 0.273, DNN: RMSE = 0.289). However, it had a larger margin of improvement in accuracy with an increase in sample size, and at $n$ = $10^6$, it reached the same level as that of NNGP. These results implied that DNN could be useful particularly in a context in which the sample size is large. In other words, in a context in which the sample size is small, its predictive accuracy does not differ much from that of OLS and this is considered to have led to the mixed findings of previous studies as discussed in Chapter 2. Figure 6 shows scatter plots depicting predicted and actual rent prices at $n$ = $10^6$.

[Figure 6: Scatter plot of predicted (horizontal axis) and actual (vertical axis) rent prices for each model (in a case of $n$ = $10^6$)], around here

From Figure 6, we can see that, across all models, the predictive accuracy is poor particularly in areas where the rent price is high. To more closely evaluate, the MAPE per logarithmic rent price range for each model is shown in Table 6. The comparison between NNGP and DNN shows that DNN was more accurate in the high-rent areas with a logarithmic rent price of 12 or greater and low-rent areas with a logarithmic rent price of 10-11. By contrast, NNGP performed better in the median-rent areas with a logarithmic rent price of 11-12.

Table 7 shows the relative frequency of the prediction error: $100\left|\frac{y_i - \hat{y}_i}{y_i}\right|$ (%) for $n = 10^6$. According to Table 7, DNN had a higher percentage of samples with an error rate of 3.5% or greater than that of NNGP (DNN: 2.461, NNGP: 2.133). Regarding the entire mean prediction result, DNN and NNGP showed similar levels of accuracy (MAPE). These results suggest that DNN shows robustness for rent price outliers but relatively high prediction errors in the vicinity of the median (rent) value. However, NNGP tends to have low predictive accuracy for samples that deviate from the median value. This would probably be because DNN is a non-linear model while NNGP is a semi-log-linear model.

These results suggest that, regarding rent price prediction models using standard explanatory variables, if the sample size is moderate ($n = 10^4$, $10^5$), Kriging (NNGP) is useful, whereas if a sufficient sample size is secured ($n = 10^6$), DNN may be promising.

[Table 6: MAPE per log (rent) range], around here

[Table 7: Relative frequency (%) ($n$=$10^6$) of prediction error rate (%)], around here

## 5. Concluding remarks

As mentioned in Chapter 1, there is a need for an accurate prediction model of real-estate sale and rent price prices for businesses and consumers. The aim of this study was to compare and discuss rent price prediction results based on regression approaches ([1] OLS and [2] spatial statistical approach (Kriging)) and [3] the machine learning approach (DNN) using various sample sizes. As the sample size increases (for example, $n = 10^5$), it is increasingly more difficult to straightforwardly apply Kriging which requires the cost of $O(n^3)$ for the inverse matrix calculation of a variance–covariance matrix. Hence as a spatial statistical approach, NNGP was used which allows application of Kriging to big data. For the machine learning approach, DNN, a representative technique, was used. DNN can consider spatial dependence through a non-linear function for position coordinates without explicitly modeling the spatial dependence as in NNGP.

For validation, from the "LIFULL HOME'S Data Set"[16], a data set for apartment rent prices in Japan—rent, lot size, location (municipality, zip code, nearest station, and walk time to nearest station), year built, room layout, building structure, and equipment for approximately 5.33 million properties across Japan—was used. To assess the effect that the sample size has on the difference in predictive accuracy, properties with missing data were eliminated and then, $n = 10^4$, $10^5$, and $10^6$ properties were completely randomly sampled to compare the rent price prediction accuracy based on approaches [1], [2], and [3]. The number of explanatory variables, $K$, was 43 including constant terms.

Our analysis showed that, with an increase in sample size, the predictive accuracy of DNN approached that of NNGP and they were nearly equal on the order of $n = 10^6$. During this experiment, standard explanatory variables that typically had been incorporated into the regression-based hedonic model were used. It is no exaggeration to say that, under these standard settings, the use of regression-based NNGP is sufficient even if the sample size is on the order of $n = 10^6$. Note, however, that DNN is expected to be useful in contexts where $K$ is even larger, e.g., when image data is used for explanatory variables. The possibility of DNN must await further investigation.

In addition, regarding both higher-end and lower-end properties whose rent prices deviate from the median, our study suggested that DNN may have a higher predictive accuracy than that of NNGP. This is because unlike NNGP, DNN can explicitly consider the non-linearity of the function form. Regarding this, the usefulness of the regression approaches that consider the non-linearity of the function form, as in the geoadditive model (Kammann and Wand, 2003), was demonstrated by the experiment conducted by Seya et al. (2011) using small samples. It will be worthwhile to test this using big data in the future.

In this study, many DNN hyper parameters were determined using optimization techniques to eliminate tailored and ad hoc setting as much as possible. Nevertheless, a certain portion of this procedure, including the setting of parameter search range, had to depend on trial and error. Because the difficulty of setting hyper parameters in DNNs poses an obstacle to their actual operation for applied researchers and practitioners who are involved in the prediction of real estate sale and rent prices, there is an urgent need to accumulate study results to resolve this issue. Additionally, it is also important to establish an effective means to set NNGP hyper parameters.

[16] https://www.nii.ac.jp/dsc/idr/lifull/homes.html

# Tables

**Table 1-1: Descriptive statistics (Continuous variables)**

| | Min | Max | Median | Mean | SD |
|---|---|---|---|---|---|
| log(rent price) (yen) | 8.57 | 20.9 | 11.1 | 11.1 | 0.402 |
| Years built (month) | 5 | 1812 | 228 | 236 | 135.6 |
| Walk time to nearest (train) station (m) | 1 | 88000 | 640 | 781.5 | 661.3 |
| Number of rooms (#) | 1 | 50 | 1 | 1.48 | 0.71 |
| Floor-area ratio (%) | 50 | 1000 | 200 | 234.1 | 130.6 |
| X (km) | -841 | 783.1 | 352.2 | 181.5 | 273.3 |
| Y (km) | 2958 | 5029 | 3931 | 3942 | 195.3 |

The "rent price" includes maintenance fees

**Table 1-2: List of explanatory variables (Discrete variables)**

| | |
|---|---|
| Direction | North, Northeast, East, Southeast, South, Southwest, West, Northwest, Other |
| Building structure | W, B, S, RC, SRC, PC, HPC, LS, ALC, RCB, Others |
| Room layout | R, K, SK, DK, SDK, LK, SLK, LDK, SLDK |
| Use district | Category □ exclusively low residential zone (1 Exc Low), Category II exclusively low residential zone (2 Exc Low), Category □ exclusively high-medium residential zone (1 Exc Med), Category II exclusively high-medium residential zone (2 Exc Med), Category I residential zone (1 Res), Category II residential zone (2 Res), Quasi-residential zone (Quasi-Res), Neighborhood commercial zone (Neighborhood Comm), Commercial zone (Commercial), Quasi-Industrial zone (Quasi-Ind), Industrial zone (Industrial), Exclusive industrial zone (Exc Ind), Others (Others) |

For building structure: W: Wooden; B: Concrete block; S: Steel frame; RC: Reinforced concrete; SRC: Steel frame reinforced concrete; PC: precast concrete; HPC: Hard precast concrete; LS: Light steel, RCB: Reinforced concrete block

For room layout: The R refers to a room where there is only one room and there is no wall to separate the bedroom from the kitchen. For the others, K: includes a kitchen; D: includes a dining room: L: includes a living room; S: additional storage room. For example, LDK is a Living, Dining, and Kitchen area.

For use district: Category I exclusively low residential zone, Category II exclusively low residential zone, Category I exclusively medium-high residential zone, Category II exclusively medium-high residential zone, Category I residential zone, Category II residential zone, Quasi-residential zone, Neighborhood commercial zone, Commercial zone, Quasi-industrial zone, Industrial zone, Exclusively industrial zone

**Table 1-3: Descriptive statistics (Discrete variables)**
**# denotes the number of cases**

| Direction | | Structure | | Use district | | Room layout | |
|---|---|---|---|---|---|---|---|
| North | 156843 | W | 1024081 | 1 Exc Low | 780638 | R | 423815 |
| Northeast | 81173 | B | 570 | 2 Exc Low | 25793 | K | 1729903 |
| East | 595252 | S | 844184 | 1 Exc Med | 689879 | SK | 6919 |
| Southeast | 473041 | RC | 1892428 | 2 Exc Med | 321441 | DK | 890584 |
| South | 1749315 | SRC | 190048 | 1 Res | 1030319 | SDK | 5123 |
| Southwest | 458125 | PC | 11924 | 2 Res | 211076 | LK | 516 |
| West | 404994 | HPS | 802 | Quasi-Res | 59863 | SLK | 138 |
| Northwest | 78836 | LS | 559974 | Neighborhood Comm | 386531 | LDK | 1505821 |
| Others | 591053 | ALC | 58373 | Commercial | 615630 | SLDK | 25813 |
| | | RCB | 597 | Quasi-Ind | 371672 | | |
| | | Others | 5651 | Industrial | 83826 | | |
| | | | | Exc Ind | 11949 | | |
| | | | | Others | 15 | | |

**Table 2: Regression analysis results using OLS (example of $n = 10^6$)**

| Variable name | Coef. | t value |
|---|---|---|
| Constant term | 10.81 | 4505 |
| Years built | -0.001155 | -444 |
| Walk time to nearest station | -0.00004840 | -98.3 |
| Floor-area ratio | 0.001294 | 228 |
| Number of rooms | 0.1486 | 257 |
| Direction_Northeast | 0.08202 | 28.3 |
| Direction_East | -0.006518 | -3.41 |
| Direction_Southeast | 0.0008989 | 0.454 |
| Direction_South | -0.02640 | -14.7 |
| Direction_Southwest | 0.001473 | 0.740 |
| Direction_West | 0.01494 | 7.49 |
| Direction_Northwest | 0.07861 | 26.9 |
| Direction_Others | -0.07103 | -36.7 |
| Structure_B | 0.2078 | 7.00 |
| Structure_S | 0.09511 | 94.9 |
| Structure_RC | 0.2418 | 274 |
| Structure_SRC | 0.3670 | 206 |
| Structure_PC | 0.2161 | 35.1 |
| Structure_HPC | 0.1186 | 5.27 |
| Structure_LS | 0.05787 | 51.5 |
| Structure_ALC | 0.09498 | 32.9 |
| Structure_RCB | 0.08847 | 3.31 |
| Structure_Others | 0.1716 | 18.9 |
| Room layout_K | 0.0414 | 35.4 |
| Room layout_SK | 0.1010 | 12.5 |
| Room layout_DK | 0.1370 | 100 |
| Room layout_SDK | 0.3696 | 39.6 |
| Room layout_LK | 0.3052 | 9.87 |
| Room layout_SLK | 0.3322 | 5.82 |
| Room layout_LDK | 0.2765 | 213 |
| Room layout_SLDK | 0.5988 | 138 |
| Use district_2 Exc Low | -0.1231 | -29.0 |
| Use district_1 Exc Med | -0.1494 | -120 |
| Use district_2 Exc Med | -0.2747 | -180 |
| Use district_1 Res | -0.2341 | -196 |
| Use district_2 Res | -0.2444 | -137 |
| Use district_ Quasi-Res | -0.2884 | -98.4 |
| Use district_ Neighborhood Comm | -0.2594 | -153 |
| Use district_ Commercial | -0.4571 | -180 |
| Use district_ Quasi-Ind | -0.1891 | -125 |
| Use district_ Industrial | -0.2451 | -97.2 |
| Use district_ Exc Ind | -0.3047 | -49.2 |
| Use district_Others | -0.3482 | -1.76 |
| Adjusted $R^2$ | 0.5178 | |

**Table 3: DNN hyper parameters and search range**

| Hyper parameters | Search range | Type |
|---|---|---|
| # of hidden layers | [1, 5] | Integer |
| # of unites | [10, 50] | Integer |
| Batch size | [32, 128] | Integer |
| # of epochs | [10, 30] | Integer |
| Learning rate | $[10^{-5}, 10^{-2}]$ (log) | Real |

**Table 4: DNN hyper parameters after optimization**

| | $n = 10^4$ | $n = 10^5$ | $n = 10^6$ |
|---|---|---|---|
| # of hidden layers | 5 | 3 | 5 |
| # of unites | [46,16,32,30,43] | [26,15,27] | [37,22,24,31,50] |
| Batch size | 33 | 45 | 125 |
| # of epochs | 15 | 19 | 25 |
| Learning rate | 0.009616156 | 0.005601793 | 0.000805416 |

**Table 5: Prediction results by sample size for each model**

| | | OLS | NNGP | DNN(Adam) | DNN(RMSprop) |
|---|---|---|---|---|---|
| $n = 10^4$ | MAE | 0.215 | 0.127 | 0.212 | 0.227 |
| | MSE | 0.074 | 0.032 | 0.083 | 0.102 |
| | RMSE | 0.273 | 0.178 | 0.289 | 0.319 |
| | MAPE | 1.938 | 1.152 | 1.920 | 2.041 |
| $n = 10^5$ | MAE | 0.216 | 0.118 | 0.155 | 0.165 |
| | MSE | 0.077 | 0.025 | 0.043 | 0.048 |
| | RMSE | 0.279 | 0.159 | 0.208 | 0.219 |
| | MAPE | 1.948 | 1.062 | 1.394 | 1.483 |
| $n = 10^6$ | MAE | 0.217 | 0.112 | 0.114 | 0.132 |
| | MSE | 0.078 | 0.024 | 0.025 | 0.033 |
| | RMSE | 0.280 | 0.155 | 0.159 | 0.182 |
| | MAPE | 1.955 | 1.013 | 1.031 | 1.195 |

**Table 6: MAPE per log (rent) range**

| Log(rent) | OLS | NNGP | DNN |
|---|---|---|---|
| ~10 | 6.787 | 3.936 | 3.940 |
| 10~10.5 | 3.456 | 1.704 | 1.508 |
| 10.5~11 | 1.657 | 0.959 | 0.931 |
| 11~11.5 | 1.602 | 0.852 | 0.920 |
| 11.5~12 | 2.712 | 1.113 | 1.328 |
| 12~12.5 | 4.718 | 1.784 | 1.734 |
| 12.5~13 | 7.549 | 3.810 | 3.265 |
| 13~ | 13.505 | 8.335 | 7.792 |

**Table 7: Relative frequency (%) ($n$=$10^6$) of prediction error rate (%)**

| Prediction error rate % | OLS | NNGP | DNN |
|---|---|---|---|
| 0 ~0.5 | 16.65 | 34.33 | 33.16 |
| 0.5 ~1.0 | 15.77 | 26.38 | 26.49 |
| 1.0 ~ 1.5 | 14.41 | 17.08 | 17.53 |
| 1.5 ~ 2.0 | 12.52 | 10.1 | 10.09 |
| 2.0 ~ 2.5 | 10.56 | 5.485 | 5.605 |
| 2.5 ~ 3.0 | 8.646 | 2.919 | 3.026 |
| 3.0 ~3.5 | 6.47 | 1.567 | 1.634 |
| 3.5 ~ | 14.98 | 2.133 | 2.461 |
| Total (%) | 100 | 100 | 100 |

**Figures**

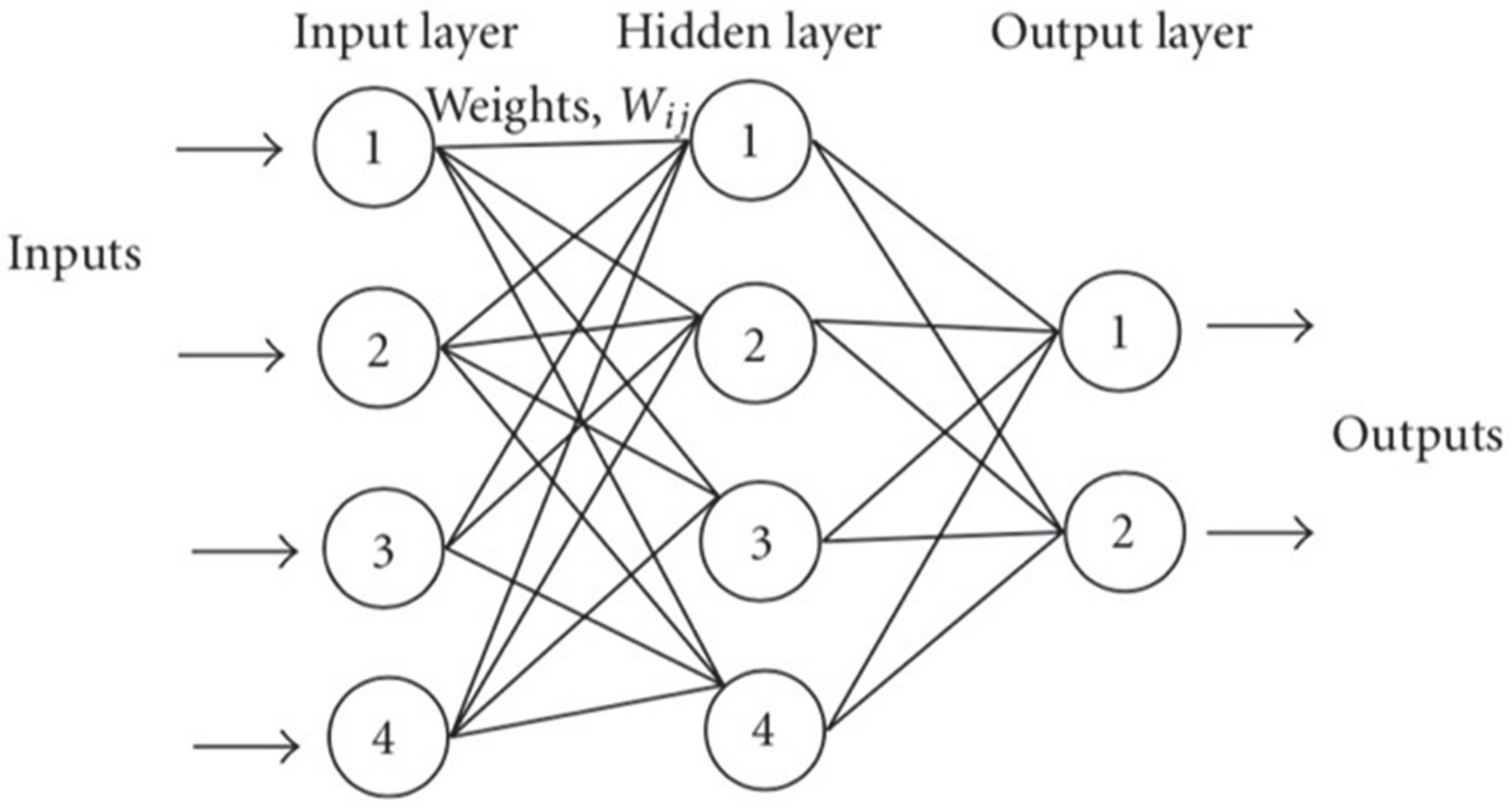

**Figure 1: Three-layered feedforward neural network**
**(Created by author in reference to Raju et al. (2011))**

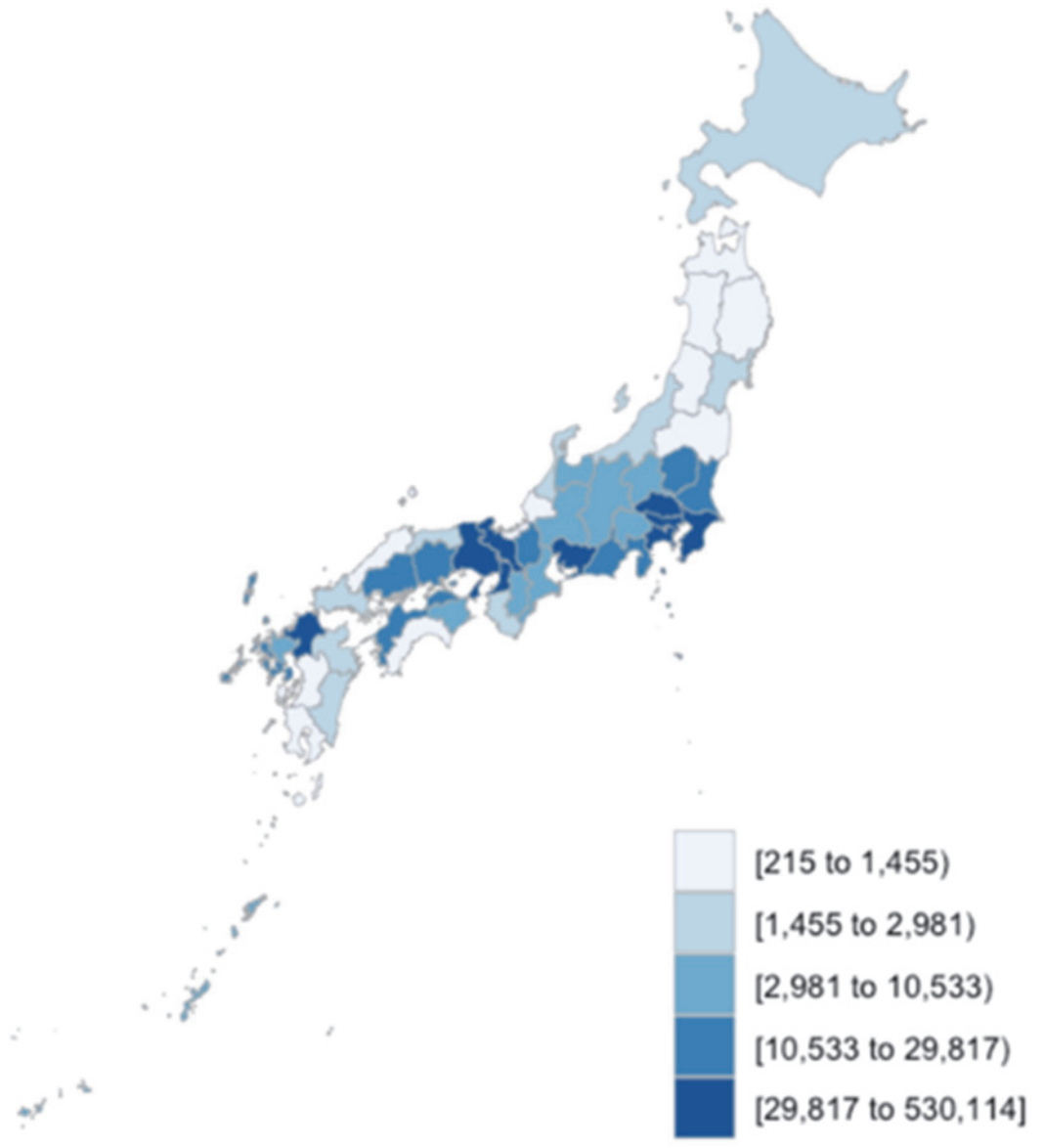

**Figure 2: Number of properties per 1000 km$^2$ for each prefecture**

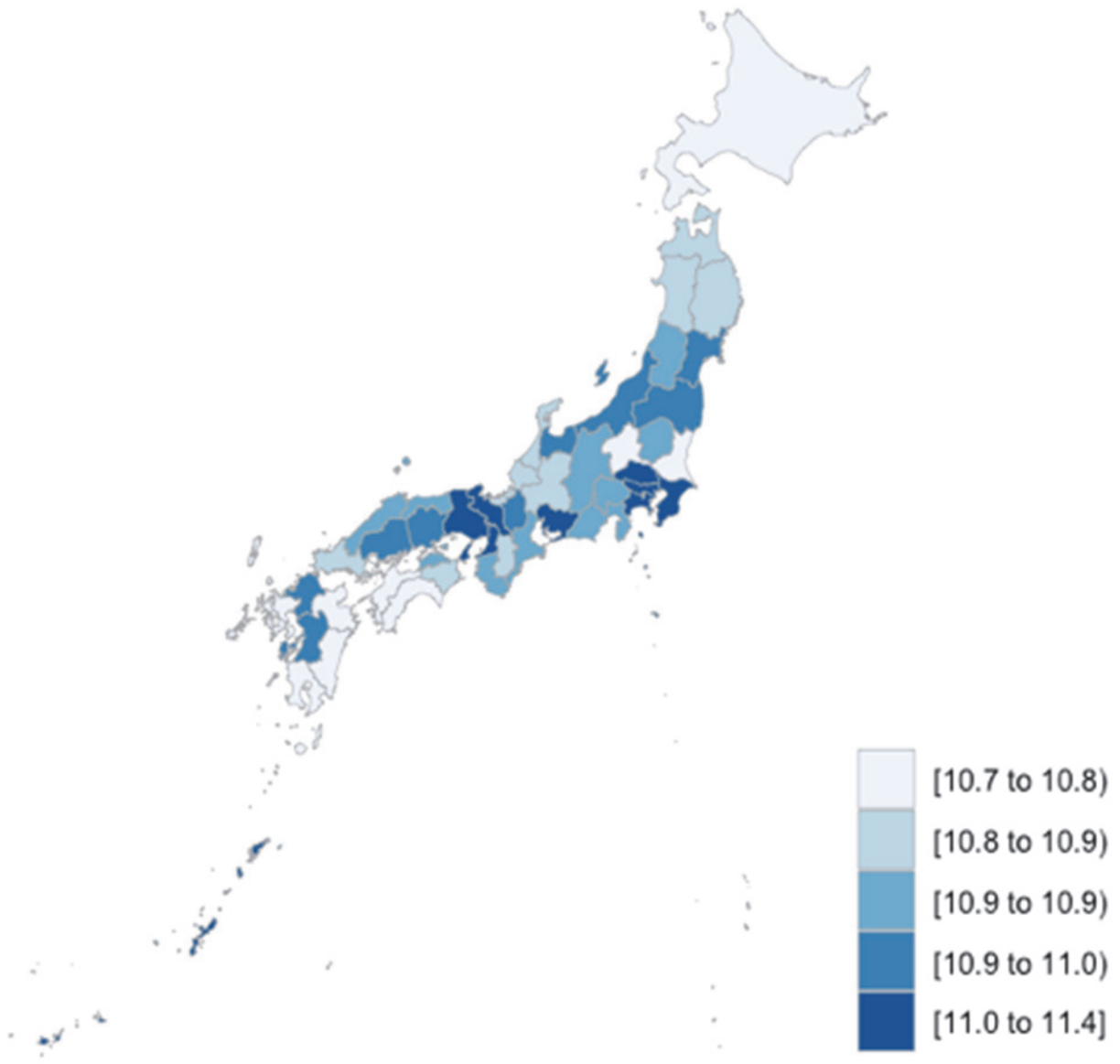

**Figure 3: log (rent prices) for each prefecture**

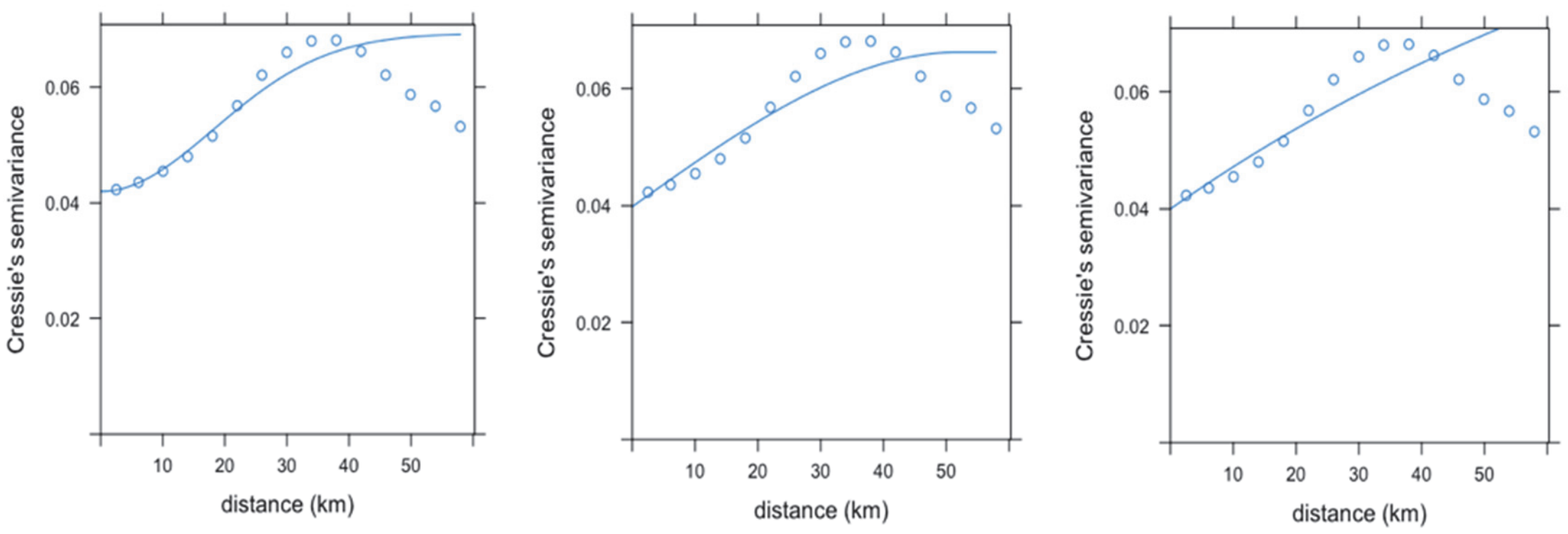

**Figure 4: Fitting of variogram functions**
**(Gaussian model; Spherical model; Exponential model)**

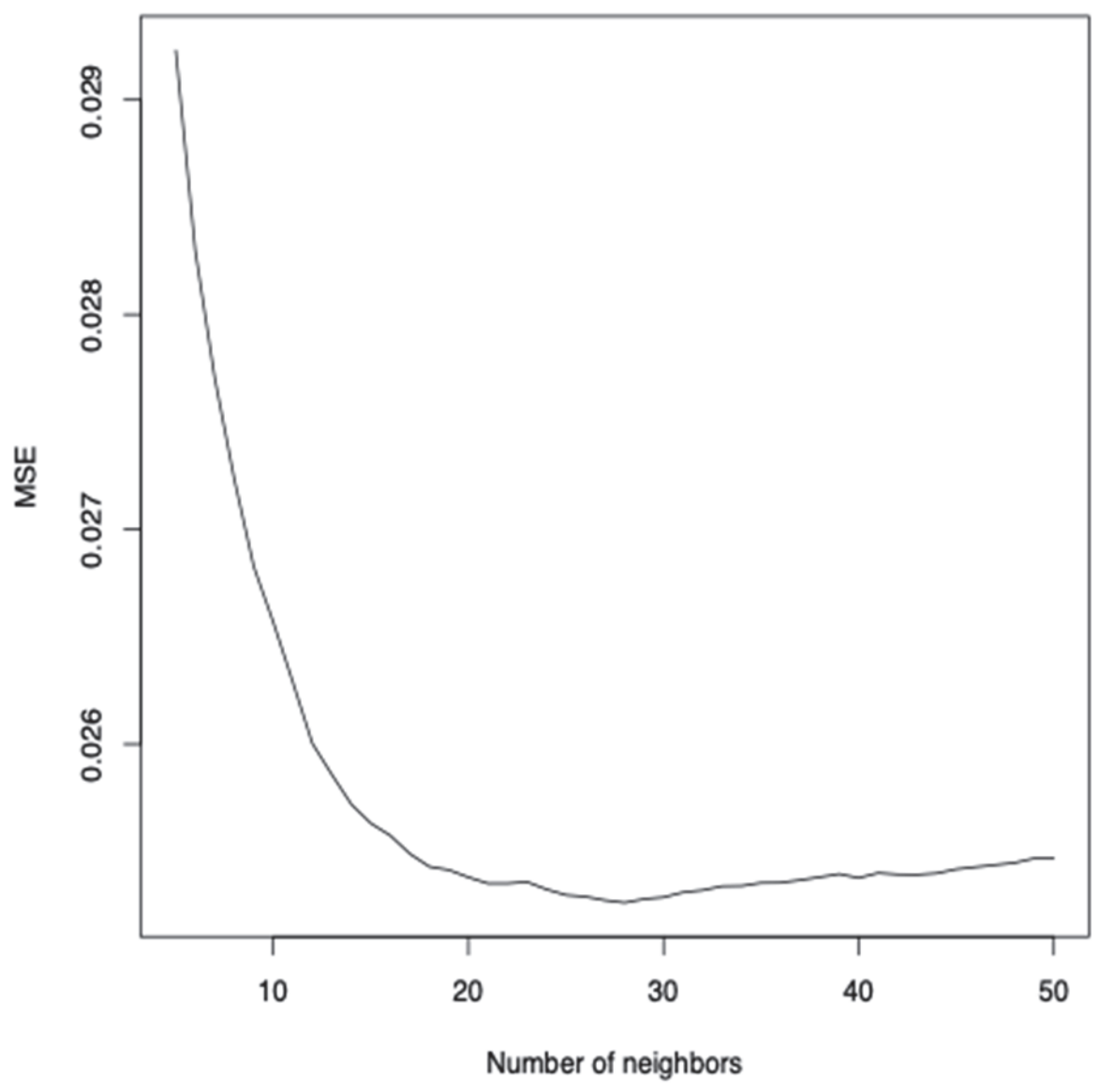

**Figure 5: Change in the MSE according to the number of nearest neighbors (in the case of $n$=$10^5$)**

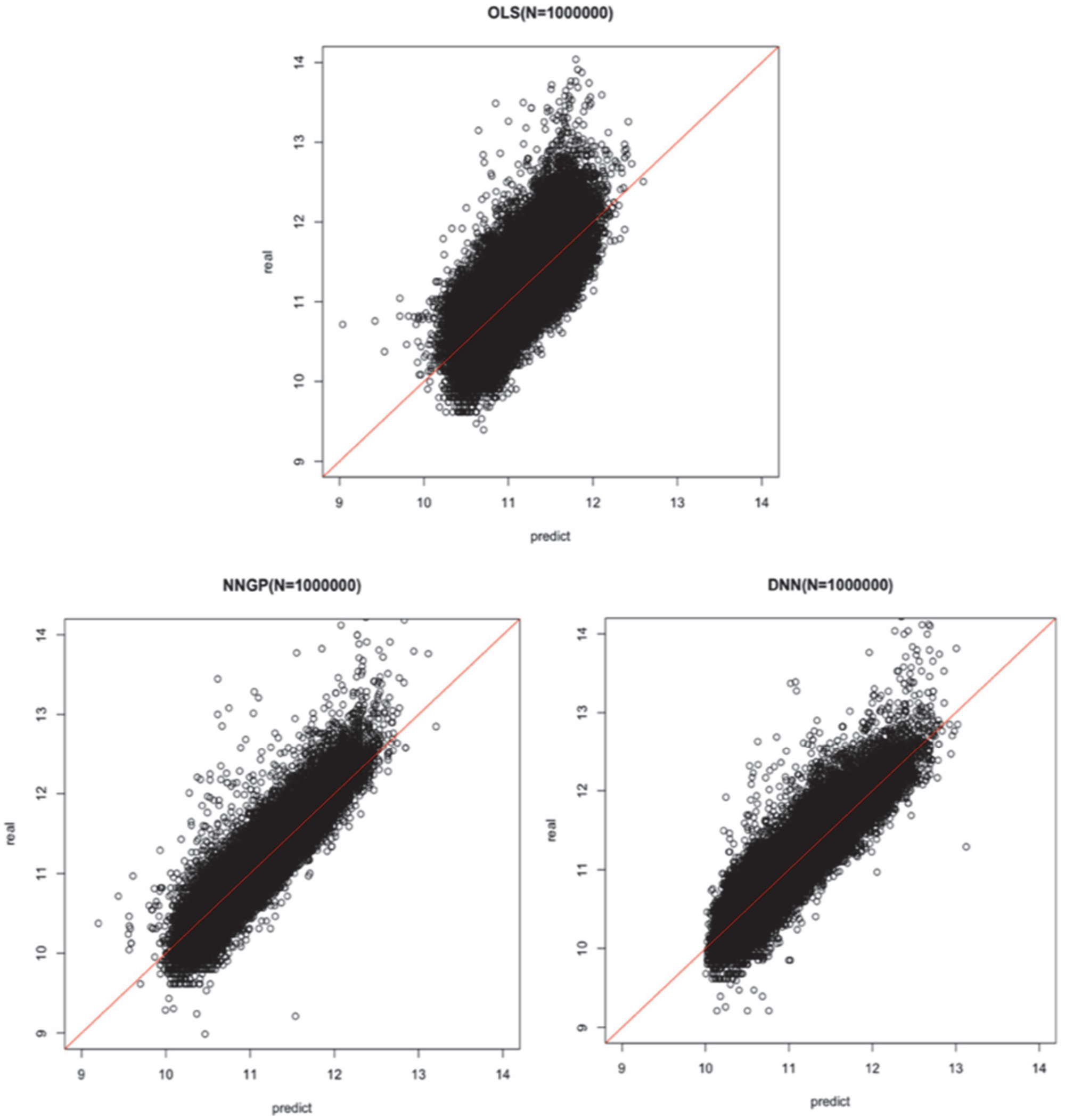

**Figure 6: Scatter plot of predicted (horizontal axis) and observed (vertical axis) rent prices for each model (in a case of $n = 10^6$)**